\documentclass[a4paper,fleqn,usenatbib]{mnras}

\usepackage{newtxtext,newtxmath}
\usepackage[T1]{fontenc}
\usepackage{ae,aecompl}

\usepackage{graphicx}	% Including figure files
\usepackage{amsmath}	% Advanced maths commands
\usepackage{amssymb}	% Extra maths symbols
\usepackage{color}
\usepackage{float}
\usepackage{courier}
\title[Positron annihilation in the nuclear outflows of the MW]{Positron Annihilation in the Nuclear Outflows of the Milky Way}

\author[F. H. Panther et al.]{
Fiona H. Panther$^{1,2}$\thanks{E-mail: fiona.panther@anu.edu.au (FHP)}, Roland M. Crocker$^{1}$, Yuval Birnboim$^{3}$, Ivo R. Seitenzahl$^{1,4}$
\newauthor
Ashley J. Ruiter$^{1,2,4}$
\\
$^{1}$Research School of Astronomy and Astrophysics, Australian National University, Canberra 2611, Australia\\
$^{2}$ARC Centre of Excellence for All-Sky Astrophysics (CAASTRO)\\
$^{3}$Racah Institute of Physics, The Hebrew University, Jerusalem, Israel\\
$^{4}$School of Physical, Environmental and Mathematical Sciences, UNSW Canberra, Australian Defence Force Academy, Canberra 2612, Australia 
}

\date{Accepted XXX. Received YYY; in original form ZZZ}

\pubyear{2017}

\begin{document}
\label{firstpage}
\pagerange{\pageref{firstpage}--\pageref{lastpage}}
\maketitle

\begin{abstract} 
Observations of soft gamma rays emanating from the Milky Way from SPI/\textit{INTEGRAL} reveal the annihilation of $\sim2\times10^{43}$ positrons every second in the Galactic bulge. The origin of these positrons, which annihilate to produce a prominent emission line centered at 511 keV, has remained mysterious since their discovery almost 50 years ago. A plausible origin for the positrons is in association with the intense star formation ongoing in the Galactic center. Moreover, there is strong evidence for a nuclear outflow in the Milky Way. We find that advective transport and subsequent annihilation of positrons in such an outflow cannot simultaneously replicate the observed morphology of positron annihilation in the Galactic bulge and satisfy the requirement that $90$ per cent of positrons annihilate once the outflow has cooled to $10^4\,\mathrm{K}$.
\end{abstract}


%%%%%%%%%%%%%%%%%%%%%%%%%%%%%%%%%%%%%%%%%%%%%%%%%%

%%%%%%%%%%%%%%%%% BODY OF PAPER %%%%%%%%%%%%%%%%%%

\section{Introduction}\label{sec:intro}
The Milky Way hosts the annihilation of $\sim$5$\times 10^{43}$ positrons each second \citep{Siegert16}. The annihilation of positrons is detected indirectly through measurements of gamma rays and is characterised by a strong emission line centered at $511\,\mathrm{keV}$, the rest mass energy of the positron (or electron). Positron annihilation in the Milky Way was first detected by balloon-bourne spectrometers in the early 1970s \citep{Johnson72}: a notable excess of emission at $\sim 0.5\,\mathrm{MeV}$ was observed to be concentrated toward the center of the Galaxy. However, the spatial resolution of such instruments was poor. The most recent observations with SPI/\textit{INTEGRAL} \citep{Knoedelseder05, Weidenspointner08, Siegert16} allow detailed morphological models of positron annihilation gamma rays to be constructed.\\
The most recent morphological models of positron annihilation in the Galactic bulge are described in \cite{Siegert16}, where the emission is modelled as the superposition of two two-dimensional gaussians. \cite{Siegert16} also describe emission from an extended thick disk,an observation highly dependent on the assumed spatial template. The more robust observation of positron annihilation in the Galactic bulge is the focus of this work, and historically the high surface brightness of positron annihilation gamma rays in this region and the high absolute positron annihilation rate - $\sim2\times10^{43}\,\mathrm{e^+\,s^{-1}}$ - have been difficult to explain. This is because most putative positrons sources are concentrated in regions of star formation in the Galactic disk \citep{Prantzos11, RKL79}, and those associated with the older stellar population of the Galactic bulge have positron yields that are currently not well constrained, such as microquasars \citep{Siegertmicroquasars}. Others employ exotic physics such as the de-excitation or annihilation of dark matter \citep[e.g.][]{Finkbeiner2007, Boehm09}.\\
The spectrum of the positron annihilation radiation observed by \textit{INTEGRAL} implies that  $\sim100$\% of positrons annihilate via interactions with neutral hydrogen \citep[charge exchange, see][for an overview]{Guessoum05}. Moreover, constraints show positrons are injected into the interstellar medium (ISM) at low energies \citep[$\leq\sim3-7\,\mathrm{MeV}$,][]{Aharonian81, Beacom06}. This likely rules out compact sources such as pulsars and millisecond pulsars as the origin of the annihilating positrons as such sources inject positrons into the ISM at energies greater than a few GeV.\\
 The observed spectrum of positron annihilation in the Galaxy is best explained with annihilation of positrons with intial energies $w_0\leq$ a few MeV in a single phase ISM with a temperature of $\sim 10^4\,\mathrm{K}$ and a degree of ionization of a few $\times 10^{-2}$ \citep{Churazov05, Siegert16} or alternatively by annihilation of these `low energy' positrons in a multiphase medium where no more than $\sim8\%$ of positrons annihilate in the very hot phase ($T\geq10^6\,\mathrm{K}$). Furthermore annihilation in the hot ($T\geq10^5\,\mathrm{K}$) or cold ($T\leq10^3\,\mathrm{K}$) media cannot make a dominant contribution to the annihilation spectrum \citep{Churazov05, Siegert16} in the presence of a multiphase medium.\\
Previous works have attempted to explain the morphology of positron annihilation gamma rays through the diffusive transport of positrons \citep{Martin2012, Alexis14} produced near the Galactic center \citep[refered to herein as `inside-out' transport, e.g. ][]{Jean09}. However, such works struggle to replicate the observed morphology of the positron annihilation signal and require the presence of a highly ordered magnetic field to aid positron diffusion out into the Galactic bulge (out to radii $\sim$kpc). In particular, \cite{Jean09} find that low energy positrons are confined to scales of $\sim200\,\mathrm{pc}$ in the presence of magnetic turbulence characterised by a Kolmogorov turbulent spectrum.\\
Diffusive transport of positrons into the Galactic bulge region was also investigated \citep{Prantzos06, Higdon09, Martin2012, Alexis14}, however the scenario presented involved positrons produced in the disk. This `outside-in' transport mechanism invokes diffusion in a similar manner to the `inside-out' diffusion scenario of \cite{Jean09}. The expected similarity of the diffusion coefficients in the Galactic disk and bulge environments\footnote{$D = 9.8 \times 10^{-4}\,\mathrm{kpc^2\,Myr^{-1}}$ for $\sim\mathrm{MeV}$ positrons in \cite{Jean09} for the Galactic bulge, as opposed to  $4.1 \times 10^{-4} \,\mathrm{kpc^2\, Myr^{-1}}$ derived in \cite{Maurin01} for the Galactic disk} suggests that diffusion in the `outside-in' transport scenario can be ruled out for the same reasons presented in the `inside-out' transport case.\\
Several works have also posited that the origin of positrons is related to our Galaxy's central supermassive black hole (SMBH) \citep[e.g.][]{Totani06, Cheng06}. In this scenario positrons are produced either in the accretion disk around the SMBH or through pair production in jets. The jet can also be invoked as a transport mechanism to distribute the positrons onto the size scales of the Galactic bulge. However, this scenario implies stringent fine-tuning constraints on the time between the launching of the jet, the total jet power and the subsequent time at which positron annihilation is observed.\\
Explaining the \textit{morphology} of positron annihilation radiation in the Galaxy represents only part of the solution to the origin of Galactic positrons. In addition, one must explain the \textit{spectrum} of positron annihilation gamma rays. In the above works, positrons are transported into neutral material where they subsequently annihilate. An alternative scenario was suggested, and found viable, by \cite{Churazov11}: positrons annihilate in an ISM that radiatively\footnote{It was noted in this work that adiabatic cooling of the ISM could be important to consider} cools from $\sim10^6\,\mathrm{K}$ to $\sim10^4\,\mathrm{K}$ on a timescale shorter than the positron lifetime in the ISM. The annihilating positrons retain no `memory' of the initial thermal conditions of the ISM. Only the ISM conditions at annihilation leave their characteristic imprint on the resulting spectrum of emitted gamma rays. The limitation of the \cite{Churazov11} model lies in the fact that the transport of positrons is not considered: positrons are assumed to annihilate in locations that replicate the morphology determined from \textit{INTEGRAL}/SPI measurements of the positron annihilation radiation described in the same work.\\
The influence of large scale motions of gas associated with star formation in the Milky Way's Galactic center have not been considered in detail with respect to resolving the origin of the Galactic bulge positrons, however positron transport in a Galactic nuclear outflow was first posited in \cite{Dermer97} in the context of observations made by OSSE/\textit{CGRO} \citep{Purcell97}. These observations revealed an extended component of positron annihilation at latitudes of up to $>15\deg$, however the observation of this feature in data from OSSE was highly model dependent \citep{Milne01b} and the feature did not appear in subsequent analysis. In the context of explaing the highly robust observation of positron annihilation in the Galactic bulge, the plausibility of such a scenario was first raised in \cite{Crocker2011}, in the context of the existence of a nuclear outflow in the Milky Way. Evidence for such an outflow emerged in \cite{BHC03}. Gamma ray structures further suggestive of such an outflow \citep[the `Fermi Bubbles';][]{Su10} were subsequently discovered in data from the \textit{FERMI} satellite. Such an outflow would cool both radiatively and adiabatically as the gas injected at the Galactic center expands \citep{Crocker15}. Moreover, it is plausible that advection could transport positrons over the $\sim\mathrm{kpc}$ distances required to explain the spatial morphology of the positron annihilation signal. Based on the work of \cite{Jean09}, the diffusion timescale for positrons to escape from a region with radius $\sim 200\,\mathrm{pc}$ is $t_\mathrm{diff} \sim 40\,\mathrm{Myr}$. This is calculated using the diffusion coefficient for MeV positrons with a Kolgomorov turbulent spectrum, derived in \cite{Jean09}. In comparison, the timescale for positrons comoving with an outflow with wind velocity $v \sim 500\,\mathrm{km\,s^{-1}}$ \citep{Crocker15} to advect to $1\,\mathrm{kpc}$ is $t_\mathrm{adv}\sim 1\,\mathrm{Myr}$, less than the diffusion timescale of the positrons. Thus, positrons can be accurately described as `frozen-in' to the plasma by magnetic turbulence and we henceforth assume they co-move with any large scale motions of this plasma.\\
In this work, we consider the transport of positrons in an analytical model of this nuclear wind and investigate whether such a scenario can replicate the two key observations of the Galactic positron annihilation signal: its morphology and the gamma ray energy spectrum.
\section{Methods}\label{sec:meth}
\subsection{Outflow Model}
We model one half of a nuclear outflow. The solid angle subtended by the outflow is fixed at $\Omega = \pi\,\mathrm{Str}$ to replicate the geometry of the Fermi Bubbles \citep{Lacki2014}, and the geometry of the outflow is that of the section of a sphere. The wind launching zone\footnote{The wind will initially accelerate, however we find that for our chosen parameter space there is no futher acceleration beyond $\sim 100\,\mathrm{pc}$} at the base of the outflow has an initial radius of $r_0 = 100\,\mathrm{pc}$ based on the radius of the Central Molecular Zone (CMZ), the region of high star fomation intensity in the Galactic center \citep{Morris96}. The evolution of the resulting wind can thus be described by the geometry of the outflow, and the mass flux, $\dot{M}$, and energy flux, $\dot{E}$, into the outflow. The initial temperature of the wind is set self-consistently with the mass and energy flux into the outflow following \cite{Strickland09}, as is the initial wind velocity at $r_0$ i.e. $T_0 = 2\mu m_p \dot{E}(5 k \dot{M})^{-1}$ and $v_0=(2\dot{E}\dot{M}^{-1})^{1/2}$. The mass density of the plasma in the outflow evolves according to mass conservation, with the initial density given by $\rho_0 =\dot{M}(2v_0\Omega r_0^2)^{-1}.$ The nuclear wind will decelerate as it does work against the gravitational potential of the Galaxy. To describe the wind deceleration we use a parameterization to approximate the gravitational potential $\Delta\phi$, described by \cite{Breitschwerdt91}: 
\begin{multline}
\phi [R, z] \simeq 1.6 \times 10^5 - \frac{8.82\times 10^4}{\sqrt{R^2+z^2+0.245}}-\\\frac{1.1\times 10^6}{\sqrt{R^2+\bigg(\sqrt{z^2+0.70}+7.26\bigg)^2}}+\frac{5.81\times10^5}{13+\sqrt{R^2+z^2}}+\\4.47\times10^4\ln\bigg({13+\sqrt{R^2+z^2}}\bigg)\,\mathrm{(km\,s^{-1})^2},
\end{multline}
where z is the height above the Galactic plane and R the Galactocentric radius, both in kpc, in cylindrical coordinates. As we assume spherical symmetry, we approximate the value of  $\Delta\phi$ by its value on the z-axis (i.e. where $R=0$ in eqn. 1). In our model, we henceforth refer to the coordinate $z$ as $r$, the radial distance from the Galactic center. The wind velocity is given as a function of the radial distance from the Galactic center r as \citep{Crocker15}: $v = v_0(1-\dot{M}\dot{E}^{-1}\Delta\phi[r])^{1/2}.$ Here $\Delta\phi[r] = \phi[r]-\phi[r_0]$.\\
The temperature of the outflow evolves through both radiative and adiabatic cooling. The radiative cooling rate is calculated assuming the plasma is in collisional ionization equilibrium (CIE). We find that in the outflow the electron-ion collision timescale \citep[e.g.][]{Lacki2014} is always shorter than the advection timescale for the outflows we describe, hence CIE is an accurate description of the cooling mode of the plasma. In addition to radiative cooling, the gas cools adiabatically as it expands, with an effective polytrope of $\gamma_\mathrm{ad} = 5/3$ for the non-relativistic plasma.\\
The ionization state of the plasma is computed from the CIE tables of \cite{MAPPINGS}. We track the evolution of the ionization fraction, and hence calculate the evolution of the neutral hydrogen number density $n_H$, ionization fraction $X_H$ and electron density $n_e$, as the energy losses of the positrons depend on the density of all species in the ISM. 
\subsection{Positron Microphysics}
We consider the propagation of positrons with initial energies $w_0<1.4\,\mathrm{MeV}$ in a nuclear outflow. This upper bound on the initial energy is consistent with a source of positrons produced in $\beta^+$ decay of radionuclides synthesised by stars and stellar end products like supernovae, and moreover is consistent with the constraints on the energies of positrons in the Galactic center from \cite{Beacom06}. For positrons emitted from $\beta^+$ unstable radionuclides (such as $^{44}$Sc and $^{56}$Co), positrons are emitted with an energy spectrum with a mean emission energy of $\sim600\,\mathrm{keV}$ and a maximum emission energy of $\sim1.5\,\mathrm{MeV}$. We calculate the energy $w_{50}$ such that 50 per cent of positrons are emitted with energies $w_{0}\leq w_{50}$. We then calculate the mean lower and upper energy at which positrons are emitted based on our calculated $w_{50}$ for positrons emitted in decay of $^{56}$Co: $w_\mathrm{low} \sim 0.4 \,\mathrm{MeV}$ and $w_\mathrm{high}\sim 0.8\,\mathrm{MeV}$ respectively. These values are close to those derived when considering positrons emitted from decay of $^{44}$Sc.\\
Our analysis does not include a detailed treatment of discrete energy loss interactions between positrons and neutral hydrogen atoms, nor do we consider the annihilation of positrons ``in-flight" (while their kinetic energies $w_0\gg 3/2 k_b T$). These processes only become important in the last few tens of years of the positrons Myr lifetimes when the positrons reach kinetic energies $w<100\,\mathrm{eV}$. The short timescales on which these processes become important, at the end of the positron lifetimes, allows us to consider only the continuous energy loss processes in calculating positron trajectories.\\
Positrons interact with all components of the ISM: neutral and ionized atoms, free electrons, the magnetic field and the radiation field. However, positrons with energies $w\ll 1\,\mathrm{GeV}$ lose the majority of their energy through ionization and plasma losses. Other radiative energy loss processes (synchrotron losses, bremsstrahlung with neutral and ionized atoms, and inverse Compton losses) are negligible for $\sim\mathrm{MeV}$ positrons \citep{Prantzos11}. The adiabatic expansion of the plasma in which the positrons are embedded mean that the positrons lose energy through adiabatic cooling. We find that the adiabatic energy losses of the positrons tend to dominate over radiative energy loss processes, and as the adiabatic energy losses of the positrons do not depend on the opening angle of the outflow, positron trajectories are not sensitive to the specific choice of opening angle. Unlike the wind fluid, the effective polytrope for the positrons evolves with the positron energy as they transition from a relativistic to non-relativistic fluid during the energy loss process. We assume an ideal equation of state (EoS), and the adiabatic index of the positrons is given by $\Gamma = 1+1/3(\beta^2/(1-(1-\beta^2)^{1/2}))$, where $\beta = (1-(1/(1+\eta)))^{1/2}$, $\eta = w/511\,\mathrm{keV}$. In reality, the EoS is dependent on the internal energy of the positrons and hence non-ideal, however we find the impact of the non-ideal EoS compared to the ideal EoS is negligible.
\subsection{Evolution of positron energy in a nuclear outflow}
In our model, we assume that positrons co-move with the expanding plasma - low energy positrons are confined by magnetic turbulence in the plasma, and their motion is thus dominated by any large-scale motions the plasma may undergo. In this scenario, positrons are transported radially outward from the wind launching zone at the wind velocity $v[t]$, embedded in an outflow that expands according to mass conservation (adiabatic expansion). Such an outflow cools as it expands.\\
We calculate the radius at which positrons are expected to annihilate. The thermalization radius, $r_\mathrm{therm}$ is the maximum radius achieved by a positron co-moving with the outflow as it loses kinetic energy from $w = w_0$ to $w_\mathrm{therm} = 6.8\,\mathrm{eV}$, corresponding to thermalization in an ISM with $T\sim10^4\,\mathrm{K}$. We choose this characteristic temperature as positrons annihilating in a plasma with this temperature will reproduce the measured linewidth of the positron annihilation line, however the exact choise for $w_\mathrm{therm}$ does not qualitatively affect our conclusions.\\
The lifetime of positrons in the outflow depends on the density of the neutral and ionized species in the ISM. As these densities vary not only with the adiabatic expansion of the plasma but also according to the radiative cooling of the plasma, the trajectories of positrons in the outflow are calculated iteratively for a range of different initial conditions. A trajectory is parameterised by the mass and energy flux into the outflow ($\dot{M}$ and $\dot{E}$ respectively) and the intial energy of the positron $w_0$. We calculate the evolution of positron energies for a grid where $\dot{M}\in\{10^{-3}, 10^0\}\,\mathrm{M_\odot\,yr^{-1}}$ and $\dot{E}\in\{10^{38}, 10^{40}\}\,\mathrm{erg\,s^{-1}}$ for positron energies $w_\mathrm{low}$ and $w_\mathrm{high}$ calculated above. The range of values chosen for the $\{\dot{E}, \dot{M}\}$ parameter space are conservatively broad estimates for the maximim and miminum values expected from nuclear star formation \citep{Crocker2012}. The initial temperature, $T_0$, and wind velocity, $v$, are defined by $\dot{E}$ and $\dot{M}$ as described above. The ionization fraction of hydrogen in the plasma is given by CIE \citep{MAPPINGS} which varies as a function of temperature. Typical inital temperatures range from $\sim 10^6 - 10^7\,\mathrm{K}$ with outflow velocities of $\sim 300 - 1500\,\mathrm{km\,s^{-1}}$. The density of the medium evolves due to mass conservation in the outflow, i.e. the density at time $t$ is
\begin{equation}
\rho[t] = \frac{\dot{M}}{2v[t]\Omega(r_0 + r)^2}, 
\end{equation}
where $t=0$ where the flow is launched at $r_0$ with velocity $v_0$ and
\begin{equation}
	r = \int_{0}^{t}v_0\bigg(1-\frac{\dot{M}}{\dot{E}}\Delta\phi[r]\bigg)^{1/2} dt
\end{equation}.\\
The temperature evolves due to adiabatic and radiative losses. The adiabatic cooling rate at time $t$ is
\begin{equation}
\frac{dT}{dt}\bigg|_\mathrm{ad}=- 2\frac{(\gamma_\mathrm{ad}-1)v[t]T_0}{(r_0+r)}\bigg(\frac{\rho[t]}{\rho_0}\bigg)^{\gamma_\mathrm{ad}-1},
\end{equation}
where $\gamma_\mathrm{ad}=5/3$ for the non-relativistic gas. 
The radiative cooling rate is
\begin{equation} 
\frac{dT}{dt}\bigg|_\mathrm{rad} = -\frac{\Lambda}{n^2}\frac{2n_\mathrm{tot}}{3k_b},
\end{equation}
where $\Lambda/n^2$ is the normalized cooling function from \cite{MAPPINGS} assuming CIE. The ionization state of the medium and the densities of the different species are computed from the same cooling tables.\\
Positrons with an inital energy of $w_0$ at $r_0$ are evolved simultaneously. Positrons lose energy through both adiabatic and radiative loss processes. As the adiabatic losses of the positrons dominate, and continuous radiative losses with species other than hydrogen in the ISM (helium and metals) are subdominant, we assume positrons interact only with hydrogen and free electrons. The adiabatic energy loss rate for positrons at time $t$ is
\begin{equation}
\frac{dw}{dt}\bigg|_\mathrm{ad}=- 2\frac{(\Gamma-1)v[t]w_0}{(r_0+r)}\bigg(\frac{\rho[t]}{\rho_0}\bigg)^{\Gamma-1}.
\end{equation}
We find that the adiabatic energy losses of the positrons always dominate over the radiative energy losses. Positrons also lose energy through ionization and Coulomb losses. Ionization losses due to interactions with neutral hydrogen are given by \citep{Ginzburg79}:
\begin{equation}
\frac{dw}{dt}\bigg|_{\mathrm{ion}} = -7.7\times10^{-9}\frac{n_{H}}{\beta}\biggl[ \ln\bigg(\frac{(\gamma-1)\gamma^2\beta^2 (511 \mathrm{keV})^2}{2I^2}\bigg)+\frac{1}{8}\biggl]\,\mathrm{eV\,s^{-1}},
\end{equation}
where $n_{H}$ is the number density of neutral hydrogen, $I = 13.8\,\mathrm{eV}$ is the ionization energy of hydrogen and $\gamma$ is the Lorentz factor for a positron with kinetic energy $w$. The energy loss rate due to coulomb scattering from the ionized ISM component are given by \citep{Huba2013},
\begin{multline}
\frac{dw}{dt}\bigg|_{\mathrm{pla}} = -1.7\times 10^{-8}\frac{n_e}{\beta}\ln{\Lambda_\mathrm{C}}\int_0^{w/k_bT} dx x^{1/2}e^{-x}-\\(w/k_bT)^{1/2}e^{-w/k_bT}) \,\mathrm{eV\,s^{-1}},
\end{multline}
where $
\Lambda_\mathrm{C} =  (k_bT/4\pi n_e e^2)^{1/2}(\mathrm{max}(2e^2/mu^2, \hbar/mu))^{-1}$ and $
u = (3E/m)^{1/2}-(8k_bT/\pi m)^{1/2}\,\mathrm{cm\,s^{-1}}$.
with $n_e$ the electron density, $m$ is the electron mass and $e$ the electron charge.\\
Trajectories of positrons are calculated and recorded until they either a) reach a maximum radius of $8\,\mathrm{kpc}$ or b) thermalize. In the latter case, we consider positrons thermalized if they reach an energy of $6.8\,\mathrm{keV}$. This positron energy is consistent with an ISM temperature of $\sim 10^4\,\mathrm{K}$. The ISM conditions (temperature, density and ionization fraction) at thermalization are recorded, as it thermalization radius, $r_\mathrm{therm}$. We then consider the regions of our input parameter space occupied by positrons that thermalize 
\section{Results}\label{sec:res}

\begin{figure}
	\includegraphics[width=\columnwidth]{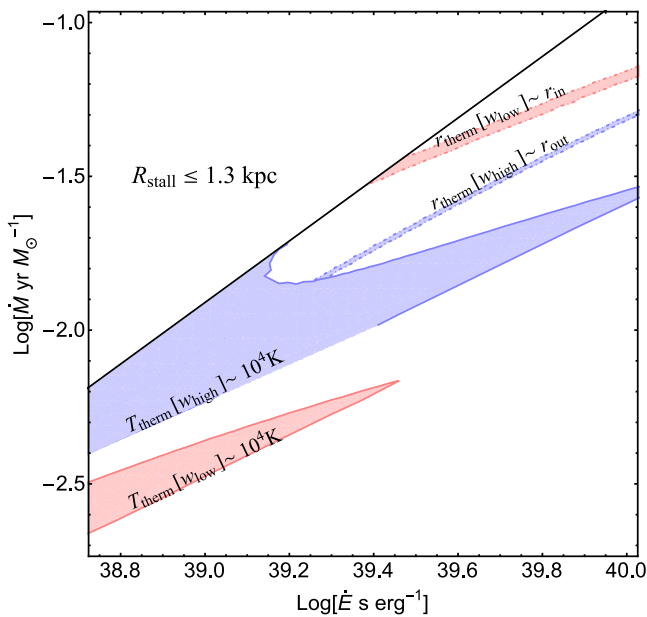}
	\caption{Contour showing regions of parameter space where $w_0 \sim 0.8\,\mathrm{MeV}$ positrons thermalize at $r_\mathrm{therm}[w_\mathrm{high}]\sim r_\mathrm{out}$ (blue shaded region, dot-dashed contours) and where the outflow has cooled to $~10^4\,\mathrm{K}$ at $r_\mathrm{out}$ (blue shaded region, solid contours). Similar contours for positrons with $w_0 \sim 0.4\,\mathrm{MeV}$ and  $r_\mathrm{therm}[w_\mathrm{high}]\sim r_\mathrm{in}$ are shown in red. The black line delineates the region of the parameter space where the outflow stalls before reaching $r_\mathrm{out}$, which are excluded in our model.}
	\label{fig:contour}
\end{figure}
For our model to replicate observations of positron annihilation in the Galactic bulge, the following must be satisfied: Firstly, the spatial distrubution of annihilating positrons must be consistent with the model described by \cite{Siegert16}, and secondly, 90 per cent of positrons must annihilate in an ISM that has cooled to $\sim 10^4\,\mathrm{K}$. To assist in comparing our results to the best fit morphology of the positron annihilation signal described by \cite{Siegert16}, we transform their two-dimensional intensity map into a one-dimensional radial intensity profile. The best fit Bulge profile from \cite{Siegert16} is the superposition of two two-dimensional Gaussian distributions representing a spatially narrow component of emission associated with the Galactic bulge, and a spatially broad component of emission associated with the same region. We calculate a mean inner radius and a mean outer radius to characterise the observed profile, i.e.
\begin{equation}
r_\mathrm{in} = \frac{\int\limits_{0}^{r_\mathrm{50}}dr I[r]r^2}{\int\limits_{0}^{r_\mathrm{50}} drI[r]r} \sim 360\,\mathrm{pc},\,r_\mathrm{out} = \frac{\int\limits_{r_\mathrm{50}}^{\infty}dr I[r]r^2}{\int\limits_{r_\mathrm{50}}^{\infty} drI[r]r} \sim 1.6\,\mathrm{kpc}
\end{equation}
where $r_\mathrm{50}$ is the root of the equation $\int^{r_\mathrm{50}}_{0}dr I[r]r/\int^{\infty}_{0}dr I[r]r = 0.5$ - radius inside which 50 per cent of the $511\,\mathrm{keV}$ flux $I[r]$ is observed. We choose to proceed with the analysis in this way - as opposed to calculating radial intensity profiles from our model - both to simplify the subsequent analysis and to take into account that the distribution of the positron annihilation radiation described in \cite{Siegert16} is a best fit model, and the smooth nature of the profile is a property of the model, not necessarily the positron annihilation signal itself.\\
The deceleration of the wind due to the gravitational potential of the Galaxy allows us to introduce a further constraint on the scenario we present. If the wind stalls before reaching $r_\mathrm{out}$, it is not possible for positrons to be transported to radii consistent with the observed intensity profile of the radiation. We calculate the regions of parameter space excluded by this constraint, which is plotted as the black solid line in fig. \ref{fig:contour}.\\
For regions of the parameter space where the wind does not stall, we calculate $r_\mathrm{therm}$ for positrons with an initial kinetic energy $w_0 = w_\mathrm{low}$ in a Galactic outflow as described in Section \ref{sec:meth}, and compare this radius to the inner characteristic radius $r_\mathrm{in}$. In Figure \ref{fig:contour}, red contours show regions of the $\{\dot{E}, \dot{M}\}$ parameter space for which positrons with $w_0 = w_\mathrm{low}$ thermalize at $r_\mathrm{in}$ (i.e. where $0.9<r_\mathrm{therm}/r_\mathrm{in}<1.1$). Overplotted, also in red, are the regions of parameter space for which the outflow cools to $8000\,\mathrm{K}\leq r_\mathrm{therm}[w_0 = w_\mathrm{low}]\leq 3\times10^4\,\mathrm{K}$ at the radius at which positrons thermalize. It is immediately obvious that these two sets of contours are disjoint, and thus there is no region of parameter space where positrons with initial energies $\sim w_\mathrm{low}$ can annihilate where the ISM temperature is consistent with the observations.\\
The solid (dot-dashed) blue contours show where positrons with initial kinetic energies $w_0 = w_\mathrm{high}$ thermalize when the ISM has cooled to $8000\,\mathrm{K}\leq r_\mathrm{therm}[w_0 = w_\mathrm{low}] \leq 3\times10^4\,\mathrm{K}$ (where positrons thermalize at $0.9<r_\mathrm{therm}/r_\mathrm{in}<1.1$). There are regions of the parameter space where positrons with initial energies $\sim w_\mathrm{high}$ annihilate in an ISM that has cooled to $\sim 10^4\,\mathrm{K}$. However, for the scenario to be presented to be consistent with observations, less than ten percent of positrons must annihilate in the hot medium. In the scenario described, the low energy positrons ($\sim50$ per cent of positrons in the outflow) annihilate before the ISM has cooled to $10^4\,\mathrm{K}$.\\
In Figure \ref{fig:contour}, one can see there is no single point in the $\{\dot{E}, \dot{M}\}$ parameter space where positrons at relevant eneriges annihilate in an ISM with a temperature of $\sim 10^4\,\mathrm{K}$ at radii consistent with the morphology of the observed extent of the $511\,\mathrm{keV}$ line. In the scenario we present, there is no region in which positrons with energies $\sim w_\mathrm{low}$ annihilate in a medium that has cooled to $10^4\,\mathrm{K}$. Consequently, the spectra of positron annihilation will be considerably broader than the observed spectrum \citep{Guessoum05} for all regions of the parameter space, even those in which positrons with energies $w_0 = w_\mathrm{high}$ thermalize and annihilate once the ISM has cooled to $\sim 10^4\,\mathrm{K}$, as the broadest component of the spectrum dominates the total observed spectrum.\\
Based on these results, we rule out a scenario where positrons are advected into the Galactic bulge by a steady state nuclear outflow as the source of Galactic bulge positrons. Our results show no sensitivity to the choice of opening angle of the described outflow. We find that the spectrum of gamma rays resulting from positron annihilation provides a stringent constraint on the annihilation sites of positrons, and our model cannot successfully reproduce the global Galactic bulge positron annihilation spectrum observed by SPI/\textit{INTEGRAL} \citep{Siegert16}.
\section{Conclusion}\label{sec:discussion}
In this work we find that positron transport in an adiabatically expanding and adiabatically and radiatively cooling steady-state outflow cannot consistently replicate either the observed morphology and spectrum of positron annihilation gamma rays. In particular, constraints on the observed spectrum of the positron annihilation radiation rule out the scenario we propose. In ruling out this scenario, where positrons are transported by large scale motions of gas from the Galactic nucleus to the Galactic bulge, we provide evidence in favour of searching for a source of Galactic positrons that is distributed in the Bulge region of the Galaxy although we do not rule out large scale diffusive transport. A distributed source such as subtypes of thermonuclear supernovae associated with old stellar populations \citep{Crocker17} or microquasars \citep{Siegertmicroquasars} can plausibly explain the observed morphology of the positron annihilation signal without invoking complex, large scale transport of positrons via diffusion.\\
\section*{Acknowledgements}
Parts of this research were conducted by the Australian Research Council Centre of Excellence for All-sky Astrophysics (CAASTRO), through project number CE110001020. IRS is supported by the Australian Research Council grant FT160100028. FHP thanks Ralph Sutherland, Geoffrey Bicknell, Dipanjan Mukherjee, Roland Diehl, Thomas Siegert, Felix Aharonian and Eugene Churazov for useful discussions. RMC thanks Geoffrey Bicknell for the calculation of the wind deceleration.

\bibliographystyle{mnras}
\bibliography{PantherBib} % if your bibtex file is called example.bib

\begin{thebibliography}{}
\makeatletter
\relax
\def\mn@urlcharsother{\let\do\@makeother \do\$\do\&\do\#\do\^\do\_\do\%\do\~}
\def\mn@doi{\begingroup\mn@urlcharsother \@ifnextchar [ {\mn@doi@}
  {\mn@doi@[]}}
\def\mn@doi@[#1]#2{\def\@tempa{#1}\ifx\@tempa\@empty \href
  {http://dx.doi.org/#2} {doi:#2}\else \href {http://dx.doi.org/#2} {#1}\fi
  \endgroup}
\def\mn@eprint#1#2{\mn@eprint@#1:#2::\@nil}
\def\mn@eprint@arXiv#1{\href {http://arxiv.org/abs/#1} {{\tt arXiv:#1}}}
\def\mn@eprint@dblp#1{\href {http://dblp.uni-trier.de/rec/bibtex/#1.xml}
  {dblp:#1}}
\def\mn@eprint@#1:#2:#3:#4\@nil{\def\@tempa {#1}\def\@tempb {#2}\def\@tempc
  {#3}\ifx \@tempc \@empty \let \@tempc \@tempb \let \@tempb \@tempa \fi \ifx
  \@tempb \@empty \def\@tempb {arXiv}\fi \@ifundefined
  {mn@eprint@\@tempb}{\@tempb:\@tempc}{\expandafter \expandafter \csname
  mn@eprint@\@tempb\endcsname \expandafter{\@tempc}}}

\bibitem[\protect\citeauthoryear{{Aharonian} \& {Atoyan}}{{Aharonian} \&
  {Atoyan}}{1981}]{Aharonian81}
{Aharonian} F.~A.,  {Atoyan} A.~M.,  1981, Pisma v Astronomicheskii Zhurnal,
  \href {http://adsabs.harvard.edu/abs/1981PAZh....7..714A} {7, 714}

\bibitem[\protect\citeauthoryear{{Alexis}, {Jean}, {Martin}  \&
  {Ferri{\`e}re}}{{Alexis} et~al.}{2014}]{Alexis14}
{Alexis} A.,  {Jean} P.,  {Martin} P.,   {Ferri{\`e}re} K.,  2014, \mn@doi
  [\aap] {10.1051/0004-6361/201322393}, \href
  {http://adsabs.harvard.edu/abs/2014A%26A...564A.108A} {564, A108}

\bibitem[\protect\citeauthoryear{{Beacom} \& {Y{\"u}ksel}}{{Beacom} \&
  {Y{\"u}ksel}}{2006}]{Beacom06}
{Beacom} J.~F.,  {Y{\"u}ksel} H.,  2006, \mn@doi [Physical Review Letters]
  {10.1103/PhysRevLett.97.071102}, \href
  {http://adsabs.harvard.edu/abs/2006PhRvL..97g1102B} {97, 071102}

\bibitem[\protect\citeauthoryear{{Bland-Hawthorn} \& {Cohen}}{{Bland-Hawthorn}
  \& {Cohen}}{2003}]{BHC03}
{Bland-Hawthorn} J.,  {Cohen} M.,  2003, \mn@doi [\apj] {10.1086/344573}, \href
  {http://adsabs.harvard.edu/abs/2003ApJ...582..246B} {582, 246}

\bibitem[\protect\citeauthoryear{{B{\oe}hm}}{{B{\oe}hm}}{2009}]{Boehm09}
{B{\oe}hm} C.,  2009, \mn@doi [New Journal of Physics]
  {10.1088/1367-2630/11/10/105009}, \href
  {http://adsabs.harvard.edu/abs/2009NJPh...11j5009B} {11, 105009}

\bibitem[\protect\citeauthoryear{{Breitschwerdt}, {McKenzie}  \&
  {Voelk}}{{Breitschwerdt} et~al.}{1991}]{Breitschwerdt91}
{Breitschwerdt} D.,  {McKenzie} J.~F.,   {Voelk} H.~J.,  1991, \aap, \href
  {http://adsabs.harvard.edu/abs/1991A%26A...245...79B} {245, 79}

\bibitem[\protect\citeauthoryear{{Cheng}, {Chernyshov}  \& {Dogiel}}{{Cheng}
  et~al.}{2006}]{Cheng06}
{Cheng} K.~S.,  {Chernyshov} D.~O.,   {Dogiel} V.~A.,  2006, \mn@doi [\apj]
  {10.1086/504583}, \href {http://adsabs.harvard.edu/abs/2006ApJ...645.1138C}
  {645, 1138}

\bibitem[\protect\citeauthoryear{{Churazov}, {Sunyaev}, {Sazonov}, {Revnivtsev}
   \& {Varshalovich}}{{Churazov} et~al.}{2005}]{Churazov05}
{Churazov} E.,  {Sunyaev} R.,  {Sazonov} S.,  {Revnivtsev} M.,   {Varshalovich}
  D.,  2005, \mn@doi [\mnras] {10.1111/j.1365-2966.2005.08757.x}, \href
  {http://adsabs.harvard.edu/abs/2005MNRAS.357.1377C} {357, 1377}

\bibitem[\protect\citeauthoryear{{Churazov}, {Sazonov}, {Tsygankov}, {Sunyaev}
  \& {Varshalovich}}{{Churazov} et~al.}{2011}]{Churazov11}
{Churazov} E.,  {Sazonov} S.,  {Tsygankov} S.,  {Sunyaev} R.,   {Varshalovich}
  D.,  2011, \mn@doi [\mnras] {10.1111/j.1365-2966.2010.17804.x}, \href
  {http://adsabs.harvard.edu/abs/2011MNRAS.411.1727C} {411, 1727}

\bibitem[\protect\citeauthoryear{{Crocker}}{{Crocker}}{2012}]{Crocker2012}
{Crocker} R.~M.,  2012, \mn@doi [\mnras] {10.1111/j.1365-2966.2012.21149.x},
  \href {http://adsabs.harvard.edu/abs/2012MNRAS.423.3512C} {423, 3512}

\bibitem[\protect\citeauthoryear{{Crocker} \& {Aharonian}}{{Crocker} \&
  {Aharonian}}{2011}]{Crocker2011}
{Crocker} R.~M.,  {Aharonian} F.,  2011, \mn@doi [Physical Review Letters]
  {10.1103/PhysRevLett.106.101102}, \href
  {http://adsabs.harvard.edu/abs/2011PhRvL.106j1102C} {106, 101102}

\bibitem[\protect\citeauthoryear{{Crocker}, {Bicknell}, {Taylor}  \&
  {Carretti}}{{Crocker} et~al.}{2015}]{Crocker15}
{Crocker} R.~M.,  {Bicknell} G.~V.,  {Taylor} A.~M.,   {Carretti} E.,  2015,
  \mn@doi [\apj] {10.1088/0004-637X/808/2/107}, \href
  {http://adsabs.harvard.edu/abs/2015ApJ...808..107C} {808, 107}

\bibitem[\protect\citeauthoryear{Crocker et~al.,}{Crocker
  et~al.}{2017}]{Crocker17}
Crocker R.~M.,  et~al., 2017, 1, 0135

\bibitem[\protect\citeauthoryear{{Dermer} \& {Skibo}}{{Dermer} \&
  {Skibo}}{1997}]{Dermer97}
{Dermer} C.~D.,  {Skibo} J.~G.,  1997, \mn@doi [\apjl] {10.1086/310870}, \href
  {http://adsabs.harvard.edu/abs/1997ApJ...487L..57D} {487, L57}

\bibitem[\protect\citeauthoryear{{Finkbeiner} \& {Weiner}}{{Finkbeiner} \&
  {Weiner}}{2007}]{Finkbeiner2007}
{Finkbeiner} D.~P.,  {Weiner} N.,  2007, \mn@doi [\prd]
  {10.1103/PhysRevD.76.083519}, \href
  {http://adsabs.harvard.edu/abs/2007PhRvD..76h3519F} {76, 083519}

\bibitem[\protect\citeauthoryear{{Ginzburg}}{{Ginzburg}}{1979}]{Ginzburg79}
{Ginzburg} V.~L.,  1979, {Theoretical physics and astrophysics}

\bibitem[\protect\citeauthoryear{{Guessoum}, {Jean}  \& {Gillard}}{{Guessoum}
  et~al.}{2005}]{Guessoum05}
{Guessoum} N.,  {Jean} P.,   {Gillard} W.,  2005, \mn@doi [\aap]
  {10.1051/0004-6361:20042454}, \href
  {http://adsabs.harvard.edu/abs/2005A%26A...436..171G} {436, 171}

\bibitem[\protect\citeauthoryear{{Higdon}, {Lingenfelter}  \&
  {Rothschild}}{{Higdon} et~al.}{2009}]{Higdon09}
{Higdon} J.~C.,  {Lingenfelter} R.~E.,   {Rothschild} R.~E.,  2009, \mn@doi
  [\apj] {10.1088/0004-637X/698/1/350}, \href
  {http://adsabs.harvard.edu/abs/2009ApJ...698..350H} {698, 350}

\bibitem[\protect\citeauthoryear{Huba}{Huba}{2013}]{Huba2013}
Huba J.~D.,  2013, {NRL PLASMA FORMULARY Supported by The Office of Naval
  Research}.
Naval Research Laboratory, Washington, DC, \url
  {http://wwwppd.nrl.navy.mil/nrlformulary/}

\bibitem[\protect\citeauthoryear{{Jean}, {Gillard}, {Marcowith}  \&
  {Ferri{\`e}re}}{{Jean} et~al.}{2009}]{Jean09}
{Jean} P.,  {Gillard} W.,  {Marcowith} A.,   {Ferri{\`e}re} K.,  2009, \mn@doi
  [\aap] {10.1051/0004-6361/200809830}, \href
  {http://adsabs.harvard.edu/abs/2009A%26A...508.1099J} {508, 1099}

\bibitem[\protect\citeauthoryear{{Johnson}, {Harnden}  \& {Haymes}}{{Johnson}
  et~al.}{1972}]{Johnson72}
{Johnson} III W.~N.,  {Harnden} Jr. F.~R.,   {Haymes} R.~C.,  1972, \mn@doi
  [\apjl] {10.1086/180878}, \href
  {http://adsabs.harvard.edu/abs/1972ApJ...172L...1J} {172, L1}

\bibitem[\protect\citeauthoryear{{Kn{\"o}dlseder} et~al.,}{{Kn{\"o}dlseder}
  et~al.}{2005}]{Knoedelseder05}
{Kn{\"o}dlseder} J.,  et~al., 2005, \mn@doi [\aap]
  {10.1051/0004-6361:20042063}, \href
  {http://adsabs.harvard.edu/abs/2005A%26A...441..513K} {441, 513}

\bibitem[\protect\citeauthoryear{{Lacki}}{{Lacki}}{2014}]{Lacki2014}
{Lacki} B.~C.,  2014, \mn@doi [\mnras] {10.1093/mnrasl/slu107}, \href
  {http://adsabs.harvard.edu/abs/2014MNRAS.444L..39L} {444, L39}

\bibitem[\protect\citeauthoryear{{Martin}, {Strong}, {Jean}, {Alexis}  \&
  {Diehl}}{{Martin} et~al.}{2012}]{Martin2012}
{Martin} P.,  {Strong} A.~W.,  {Jean} P.,  {Alexis} A.,   {Diehl} R.,  2012,
  \mn@doi [\aap] {10.1051/0004-6361/201118721}, \href
  {http://adsabs.harvard.edu/abs/2012A%26A...543A...3M} {543, A3}

\bibitem[\protect\citeauthoryear{{Maurin}, {Donato}, {Taillet}  \&
  {Salati}}{{Maurin} et~al.}{2001}]{Maurin01}
{Maurin} D.,  {Donato} F.,  {Taillet} R.,   {Salati} P.,  2001, \mn@doi [\apj]
  {10.1086/321496}, \href {http://adsabs.harvard.edu/abs/2001ApJ...555..585M}
  {555, 585}

\bibitem[\protect\citeauthoryear{{Milne}, {Kurfess}, {Kinzer}  \&
  {Leising}}{{Milne} et~al.}{2001}]{Milne01b}
{Milne} P.~A.,  {Kurfess} J.~D.,  {Kinzer} R.~L.,   {Leising} M.~D.,  2001, in
  {Ritz} S.,  {Gehrels} N.,   {Shrader} C.~R.,  eds,  American Institute of
  Physics Conference Series Vol. 587, Gamma 2001: Gamma-Ray Astrophysics. pp
  11--15 (\mn@eprint {} {astro-ph/0106157}), \mn@doi{10.1063/1.1419363}

\bibitem[\protect\citeauthoryear{{Morris} \& {Serabyn}}{{Morris} \&
  {Serabyn}}{1996}]{Morris96}
{Morris} M.,  {Serabyn} E.,  1996, \mn@doi [\araa]
  {10.1146/annurev.astro.34.1.645}, \href
  {http://adsabs.harvard.edu/abs/1996ARA%26A..34..645M} {34, 645}

\bibitem[\protect\citeauthoryear{{Prantzos}}{{Prantzos}}{2006}]{Prantzos06}
{Prantzos} N.,  2006, \mn@doi [\aap] {10.1051/0004-6361:20052811}, \href
  {http://adsabs.harvard.edu/abs/2006A%26A...449..869P} {449, 869}

\bibitem[\protect\citeauthoryear{{Prantzos} et~al.,}{{Prantzos}
  et~al.}{2011}]{Prantzos11}
{Prantzos} N.,  et~al., 2011, \mn@doi [Reviews of Modern Physics]
  {10.1103/RevModPhys.83.1001}, \href
  {http://adsabs.harvard.edu/abs/2011RvMP...83.1001P} {83, 1001}

\bibitem[\protect\citeauthoryear{{Purcell} et~al.,}{{Purcell}
  et~al.}{1997}]{Purcell97}
{Purcell} W.~R.,  et~al., 1997, \mn@doi [\apj] {10.1086/304994}, \href
  {http://adsabs.harvard.edu/abs/1997ApJ...491..725P} {491, 725}

\bibitem[\protect\citeauthoryear{{Ramaty}, {Kozlovsky}  \&
  {Lingenfelter}}{{Ramaty} et~al.}{1979}]{RKL79}
{Ramaty} R.,  {Kozlovsky} B.,   {Lingenfelter} R.~E.,  1979, \mn@doi [\apjs]
  {10.1086/190596}, \href {http://adsabs.harvard.edu/abs/1979ApJS...40..487R}
  {40, 487}

\bibitem[\protect\citeauthoryear{{Siegert} et~al.,}{{Siegert}
  et~al.}{2016a}]{Siegertmicroquasars}
{Siegert} T.,  et~al., 2016a, \mn@doi [\nat] {10.1038/nature16978}, \href
  {http://adsabs.harvard.edu/abs/2016Natur.531..341S} {531, 341}

\bibitem[\protect\citeauthoryear{{Siegert}, {Diehl}, {Khachatryan}, {Krause},
  {Guglielmetti}, {Greiner}, {Strong}  \& {Zhang}}{{Siegert}
  et~al.}{2016b}]{Siegert16}
{Siegert} T.,  {Diehl} R.,  {Khachatryan} G.,  {Krause} M.~G.~H.,
  {Guglielmetti} F.,  {Greiner} J.,  {Strong} A.~W.,   {Zhang} X.,  2016b,
  \mn@doi [\aap] {10.1051/0004-6361/201527510}, \href
  {http://adsabs.harvard.edu/abs/2016A%26A...586A..84S} {586, A84}

\bibitem[\protect\citeauthoryear{{Strickland} \& {Heckman}}{{Strickland} \&
  {Heckman}}{2009}]{Strickland09}
{Strickland} D.~K.,  {Heckman} T.~M.,  2009, \mn@doi [\apj]
  {10.1088/0004-637X/697/2/2030}, \href
  {http://adsabs.harvard.edu/abs/2009ApJ...697.2030S} {697, 2030}

\bibitem[\protect\citeauthoryear{{Su}, {Slatyer}  \& {Finkbeiner}}{{Su}
  et~al.}{2010}]{Su10}
{Su} M.,  {Slatyer} T.~R.,   {Finkbeiner} D.~P.,  2010, \mn@doi [\apj]
  {10.1088/0004-637X/724/2/1044}, \href
  {http://adsabs.harvard.edu/abs/2010ApJ...724.1044S} {724, 1044}

\bibitem[\protect\citeauthoryear{{Sutherland}, {Dopita}, {Binette}  \&
  {Groves}}{{Sutherland} et~al.}{2013}]{MAPPINGS}
{Sutherland} R.,  {Dopita} M.,  {Binette} L.,   {Groves} B.,  2013, {MAPPINGS
  III: Modelling And Prediction in PhotoIonized Nebulae and Gasdynamical
  Shocks}, Astrophysics Source Code Library (\mn@eprint {ascl} {1306.008})

\bibitem[\protect\citeauthoryear{{Totani}}{{Totani}}{2006}]{Totani06}
{Totani} T.,  2006, \mn@doi [\pasj] {10.1093/pasj/58.6.965}, \href
  {http://adsabs.harvard.edu/abs/2006PASJ...58..965T} {58, 965}

\bibitem[\protect\citeauthoryear{{Weidenspointner} et~al.,}{{Weidenspointner}
  et~al.}{2008}]{Weidenspointner08}
{Weidenspointner} G.,  et~al., 2008, \mn@doi [\nat] {10.1038/nature06490},
  \href {http://adsabs.harvard.edu/abs/2008Natur.451..159W} {451, 159}

\makeatother
\end{thebibliography}

%%%%%%%%%%%%%%%%%%%%%%%%%%%%%%%%%%%%%%%%%%%%%%%%%%

%%%%%%%%%%%%%%%%% APPENDICES %%%%%%%%%%%%%%%%%%%%%
%
%\appendix

%%%%%%%%%%%%%%%%%%%%%%%%%%%%%%%%%%%%%%%%%%%%%%%%%%

% Don't change these lines
\bsp	% typesetting comment
\label{lastpage}
\end{document}